\documentclass[aps,prl,showpacs,twocolumn]{revtex4}
\usepackage{bm}
\usepackage{graphicx,xr}

\newcommand{\nvec}[1]{{\mathbf #1}}

\begin{document}

\title[Effective interaction for sheet arrangements]{ Transverse dipole-dipole effective interaction for sheet arrangements}

\author{Ladislav Kocbach} 
\email{ladislav.kocbach@ift.uib.no}
\author{Suhail Lubbad}
\email{suhail.lubbad@gmail.com}
\affiliation{Department of Physics and Technology, University of Bergen, All\'egaten
  55, 5007 Bergen, Norway}

\begin{abstract}
We have succeeded to develop a model pair interaction which when added to
a system of interacting particles  can be tuned to arrange
the interacting objects into sheets. The interaction is
based on the decomposition of the dipole-dipole interaction
into two components, one parallel and one perpendicular to the connecting line
between the dipoles, and keeping only perpendicular part here. Various aspects
of this simple interaction are discussed, in particular in connection to
two recent papers on self assembly of carbon nanostructures. On the other hand,
the features discussed are quite general and might be of interest in
different areas of microscopic modeling.

\end{abstract}

\pacs{36.20.Hb, 63.22.Np, 81.16.Rf, 87.85.Qr}

\maketitle

In many areas of physics there is often a need for simple
effective interactions. Perhaps the most famous might be
the Lennard-Jones potential or  the spin-spin interaction of the
Heisenberg ferromagnet, two models which are well known 
and used far outside
the original applications. Our aim has been to understand the
emergence of various
geometrical arrangements and
we have succeeded to develop a model two-body interaction which when added to
a system of interacting particles would  arrange them into sheets instead
of the expected three-dimensional structures.
This model interaction can primarily be of use in atomistic simulations,
but possibly also for  systems consisting of more complex particles,
in principle at any
scale.

We start from the usual form of the dipole-dipole interaction
\begin{equation}
   U(\nvec{r}_{12}) =  \frac{A}{r_{12}^{\ \ 3}}  \left[ \nvec{m}_1
   \cdot \nvec{m}_2 - 3 \left( \nvec{m}_1 \cdot \nvec{e}_{12}
   \right) \left( \nvec{m}_2 \cdot \nvec{e}_{12} \right)
   \right] 
   \label{dipole_std}
\end{equation}   
where $\nvec{r}_{12}={r}_{12} \nvec{e}_{12}$ is the vector connecting the two dipoles
and $ \nvec{e}_{12}$ its unit vector.
It has probably been noticed by many, but apparently not discussed 
by anybody in printed form,    
that eq. \ref{dipole_std}
can be rewritten as
\begin{equation}
 U =  f(r_{12}) 
\left[ 
   \nvec{m}_{1}^\perp \cdot \nvec{m}_{2}^\perp 
   - 2   \nvec{m}_{1}^\|    \cdot \nvec{m}_{2}^\|
   \right] 
   \label{dipoletrafo}
\end{equation}   
The notation used follows from decomposition of both  dipoles into two components
$$
 m_1 = m_{1}^\|  \nvec{e}_{12} + \nvec{m}_{1}^\perp 
   = 
   \nvec{m}_{1}^\| +  \nvec{m}_{1}^\perp
$$   
\begin{figure}[htb]
\center{
\includegraphics[height=5.5cm]{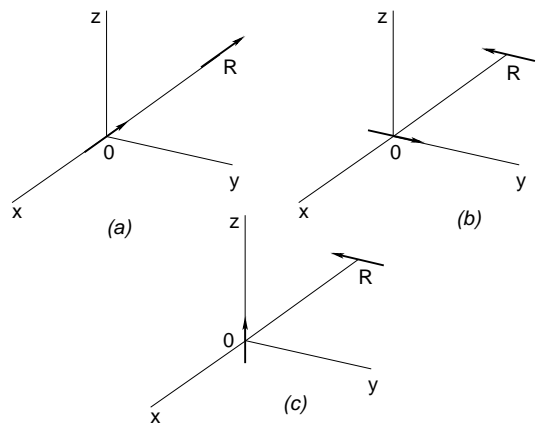}
}
 \caption{
 The dipoles attracting each other when 
 positive maximum scalar product (a) due to the $-2$ term, 
 and when the scalar product is negative (b) due to the first term.
 There is a zero potential and thus zero force (c) when the
 dipols are forced to remain perpendicular.
\label{lab.fig.general} }
 %%%
\end{figure}
The figure \ref{lab.fig.general} reminds us of the situations when each of the terms dominate. 
For the real dipoles
the two terms are added exactly in the given radial form, but effective 
interactions may naturally assume any radial form
and any relative strength of the two terms suitable for the physical model in question.
Our model consists of only the part containing the perpendicular terms, or 
alignment of the intrinsic vectors in direction perpendicular to the
connecting line of the two objects.  
We propose to call this model interaction "transverse 
dipole-dipole interaction", or TDDI. The effective pair potential
is chosen as
\begin{equation}
 U \left( {r}_{ij}, \nvec{e}_{ij},  \nvec{m}_i , \nvec{m}_j             \right)
 =
        g( r_{ij}) 
       G( 1 - \left|      \nvec{m}_{i}^\perp \cdot \nvec{m}_{j}^\perp \right|   )
   \label{dipoleeffective}
\end{equation}   
where all the vectors $ \nvec{e}_{ij}$,  $\nvec{m}_i$ , $\nvec{m}_j $
are unit vectors and the function $ g( r_{ij}) $
is an arbitrary function suitable for the model system in question.
It can be of short range or long range, but preferably reaching at least two
nearest neighbours to define the plane. 
The argument of function $G(u)$ is zero when the two vectors point in parallel
lines, in both directions. Thus  $G(u)$ must be sufficiently strongly peaked 
at $u=0$.

Considering first just four particles with such interaction, 
this term will be zero only 
when all four particles are in the same plane, having thus a minimum.
If there is also some other additional interaction leading to 
equilibrium distances, 
both the discussed four particle as well as more particle systems
would prefer  sheets or single layers, much like the graphene sheets, as long
as the range of $ g( r_{12}) $ covers at least two neighbours. 
If there is an interaction
arranging the objects at certain distances and perhaps also with well defined
bonding angles, but without the ability to assure a planar arrangement, adding a fictive
model vector - transverse dipole (TD) - to every particle, the TDDI would lead to sheet-like
behavior, even if it is only a two-body interaction. Since parallel TDs are preferred,
any three neighbours would align them perpendicular to the plane in which they are lying.
When all these planes become one plane, the TDDI is minimized. Thus TDDI added to
a suitable two or three-body interaction would lead to self assembly of sheets.

Combining TDDI with additional terms is important because 
when TDDI would be the only interaction, its radial dependence $g(r)$ 
would need to have a minimum to assure some equilibrium distances. 
This could be in conflict with the aim
to establish a stable sheet structure, since it 
could lead to a close packing situation. 
All the effective 
internal vectors could be  partially aligned but a closed packed 
structure will result from the existence of the minimum in the function
 $g(r)$. Thus the discussed model interaction should only be used in addition
 to other inter-particle forces. In general, the  $g(r)$ function 
 should be monotonically decreasing with distance, in most cases simply 
 providing a cut-off.

The proposed TDDI is a pair interaction. The limitation to the perpendicular
term for a single pair leads only to reorientation of the two
internal spins with no influence on the pair of particles. The same is true
for a group of three particles. All their connecting lines are in one plane,
so that the internal vectors can reorient themselves without any 
influence on the positions of the particles. When a fourth particle is added, 
the torque on the internal vectors will start forcing the particles into one plane.
\begin{figure}[htb]
\center{
\includegraphics[height=2.5cm]{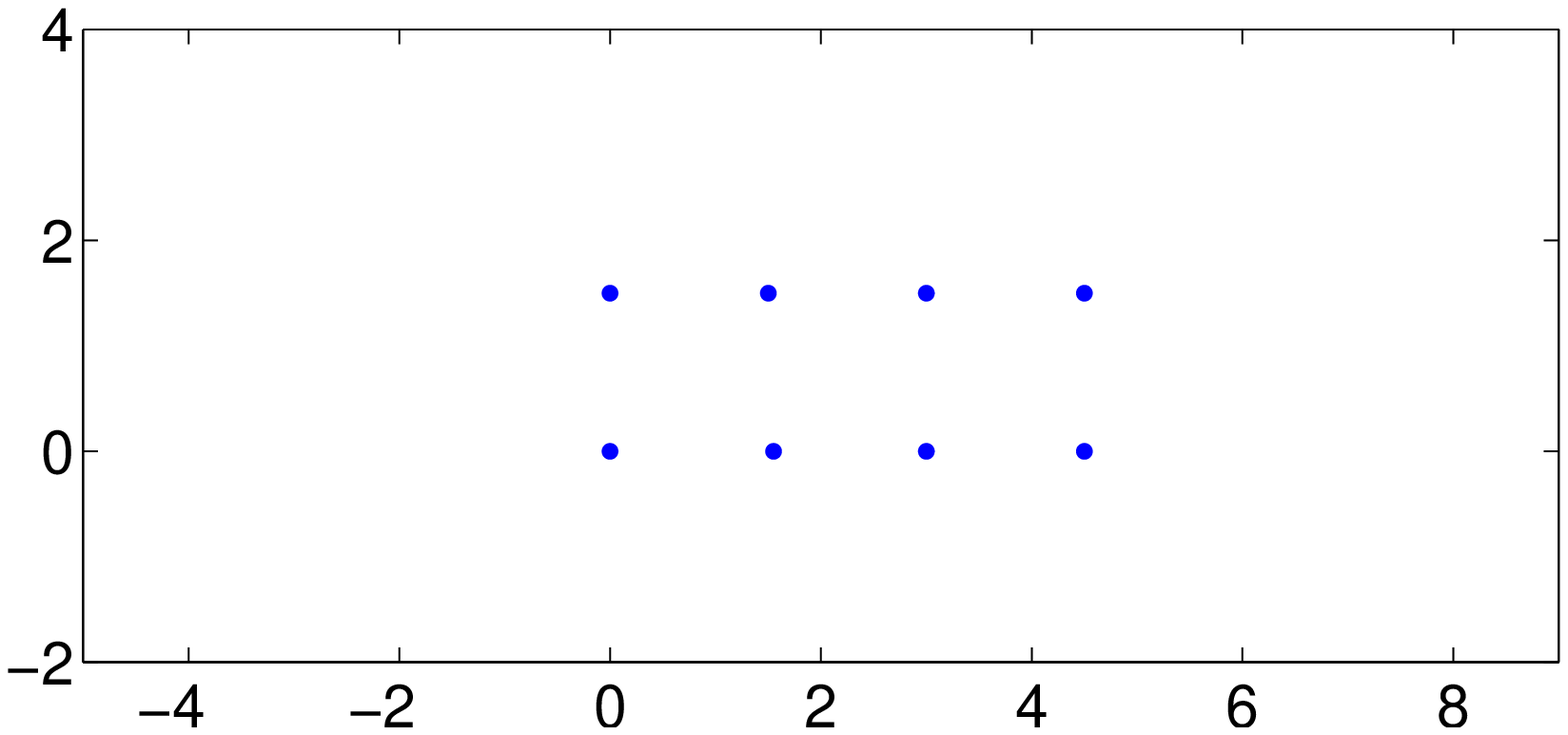} \\
\includegraphics[height=2.5cm]{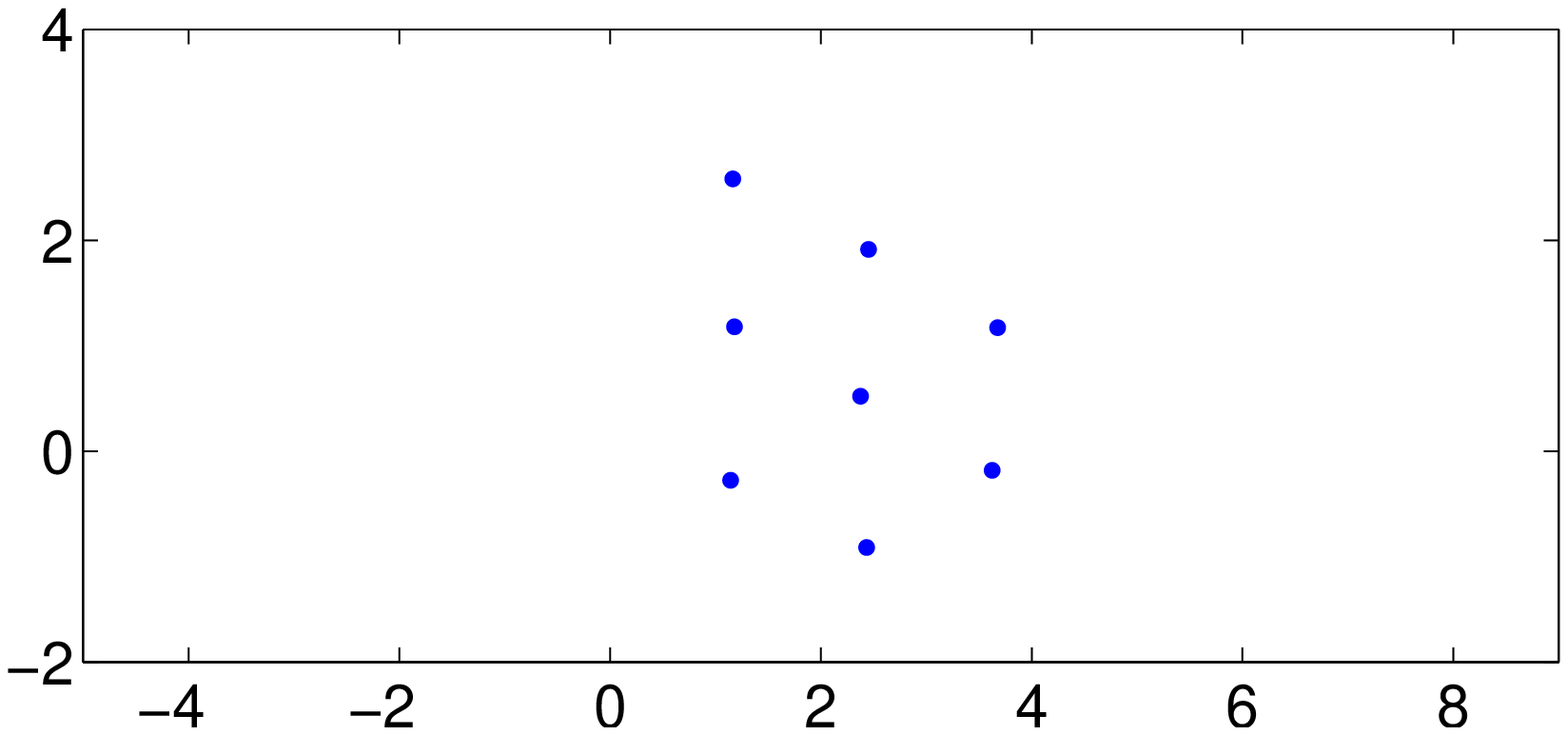}
}
 \caption{
Eight particles, Morse interaction. Above: starting configuration Below: after some time, with friction
\label{no_spin_fig} }
 %%%
\end{figure}

The present paper is to some degree inspired 
by the work of
Rechtsman, Stillinger and Torquato 
 \cite{stillinger2007}, who show 
how to build structures of diamond and wurtzite type
using special pair interactions.
These {\it isotropic} pair interactions have minima where
the  radial distribution functions
(RDF) of the target structures have peaks.
The principle proposed here is somewhat similar.
\begin{figure}[htb]
\center{
\includegraphics[height=2.5cm]{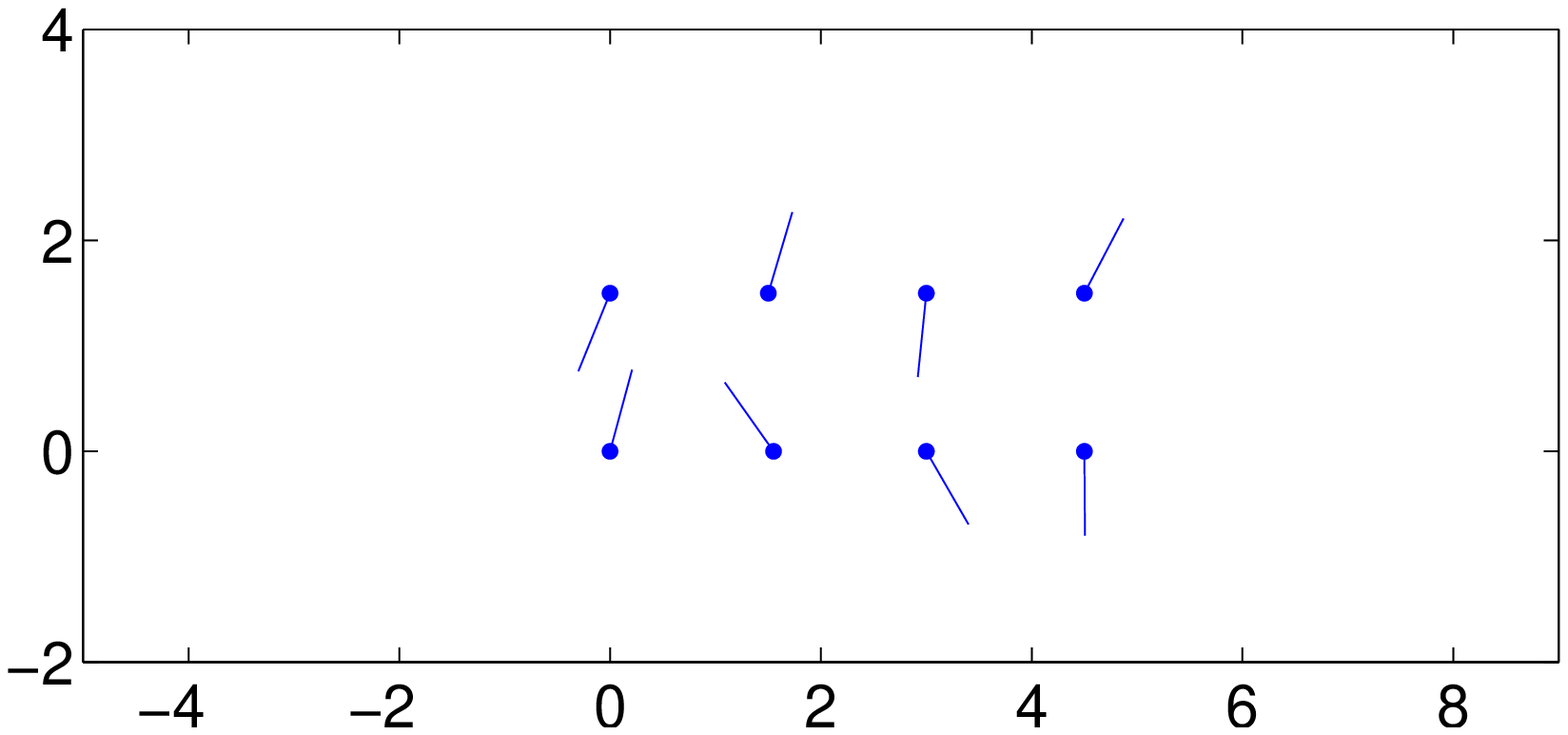} \\
\includegraphics[height=2.5cm]{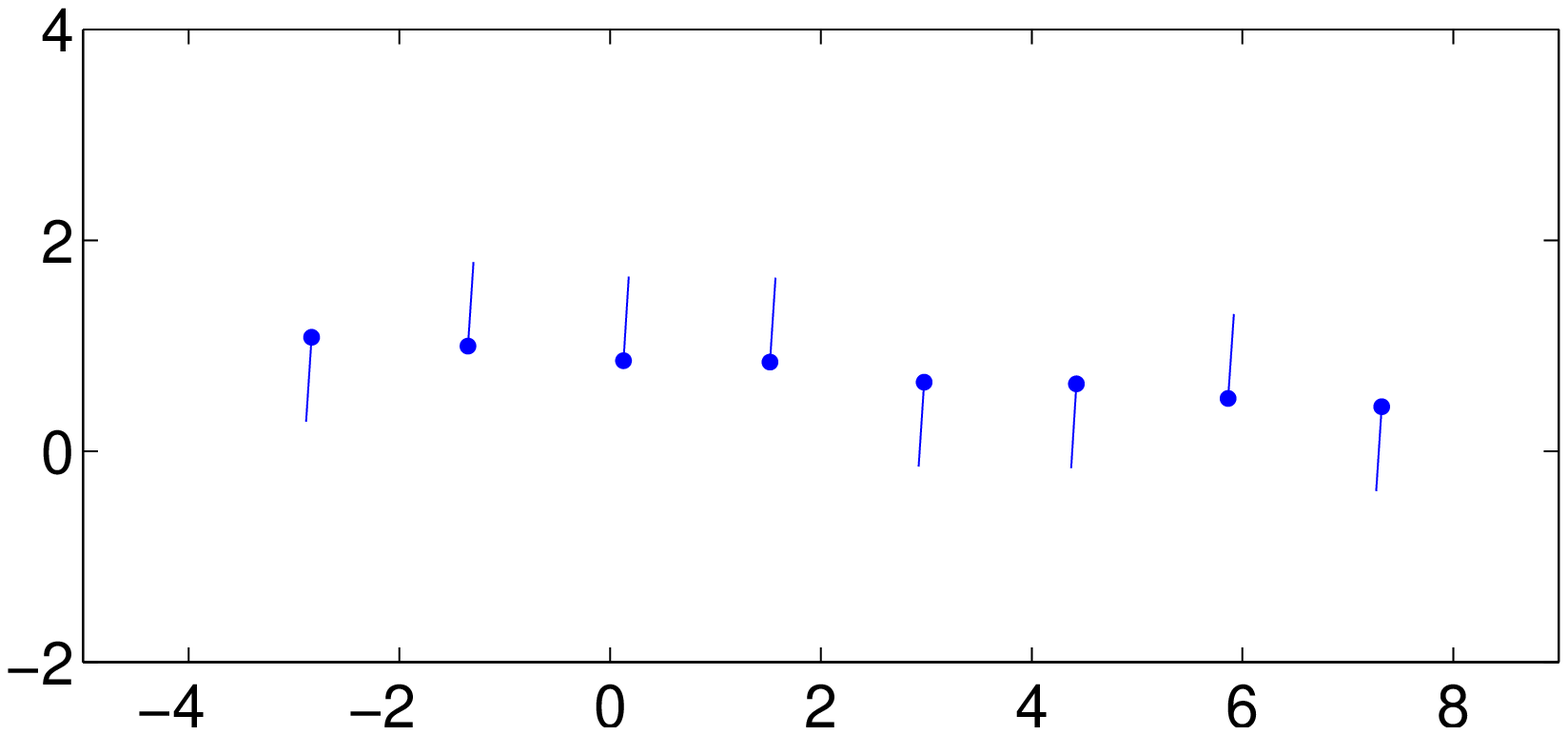}
}
 \caption{Eight particles as in fig. \ref{no_spin_fig}, but now with the TDDI forces added to the same
 Morse interaction. Above: starting configuration, the same as before; Below: after some time, with friction
\label{the_spin_fig} }
 %%%
\end{figure}

Tewary and   Yang in ref \cite{graphene_tersoff} construct 
a parametric potential for a graphene sheet. 
They wish to achive similar effect as described here, but
they simulate it by combining three body forces in a rather
complicated way. They use a rather complex 
many body potential of Tersoff type \cite{Tersoff_1988_PRB}
with an addition of three-body terms adjusting the planar geometry.
Our proposed intrinsic vector model is 
much simpler and more flexible, and computationally appears as 
a two-body term.
We suggest that a much simpler model than the one
of ref. \cite{graphene_tersoff} could be based on adjusting 
the Stillinger and Weber model \cite{stillinger1985}, which is a 
two plus three body interaction for tetrahedral diamond
correlations, to the plane geometry with the addition of 
terms of the type proposed here (see also ref. \cite{obmd})

The effect of the TDDI interaction can be analyzed and explored in
two dimensional case much easier
than in three dimensions. 
In two dimensional case the role of the sheet of particles
is played by a string of particles. We have implemented the TDDI 
into a small program which solves the molecular dynamics-like problem
in two dimensions. The interactions are Morse potential, which is known to 
lead to the triangular-hexagonal closed packed arrangement 
(isotropic forces with an equilibrium distance). Then we added the 
TDDI interaction. Results of two runs are shown in figures
\ref{no_spin_fig} and \ref{the_spin_fig}. The variant of TDDI we used here
is
\begin{equation}
 U \left(  \nvec{r}_{ij},  \nvec{m}_i , \nvec{m}_j             \right)
 =
    U_0
     \left[    1 - \left(      \nvec{m}_{i}^\perp \cdot \nvec{m}_{j}^\perp \right)^2   
     \right]          .
   \label{dipole_used}
\end{equation}   
For the small number of particles the motion becomes quite disordered, 
therefore we have added a friction term to be able to see the "cooled" structures.
Note that the starting situation is a regular, but with one particle slightly displaced.
Without any tiny displacement a regular structure will simply "breathe".
The difference between the calculations is clear and demonstrate the effect.
The physical dipole-dipole interaction where the first term dominates
would not lead to any similar effect, but it would also lead to some anisotropy.

The forces on the particles are given by the gradients of the
potential energy, and it is instructive to inspect the resulting formulae.
The gradient with respect to the $\nvec{r}_i$ coordinate of the transverse TDDI term
$ \nvec{m}^\perp_i  \cdot    \nvec{m}^\perp_j$, is obtained after some 
rather elementary algebra to be
\begin{equation}
\nabla_{r_i}  \left(  
 \nvec{m}^\perp_i  \cdot    \nvec{m}^\perp_j   
 \right)    
 =
 -  \frac{1}{r_{ij}}  
    \left(
           m_{i}^\|   \nvec{m}^\perp_j
           +
           m_{j}^\|  \nvec{m}^\perp_i 
    \right)
   \label{transv_gradient}
\end{equation}   
Clearly, this is the part important for the motion of the 
particles when the vectors $ \nvec{m}_i $, $   \nvec{m}_j$, are
"frozen", or kept constant,  the torque terms
leading to their reorientation will be addressed below. The above gradient must be used in the 
evaluation of the forces on each of the particles. For a pair potential
\begin{equation}
W( \nvec{r}_{ i},  \nvec{m}_{ i}, \nvec{r}_{ j}, \nvec{m}_{ j}) =
V(r_{ i j}) 
     +   G_0 \     
 \left[     
  1 -      \left(     \nvec{m}^\perp_i  \cdot    \nvec{m}^\perp_j  \right) ^2
 \right] 
   \label{model_potential}
\end{equation}   
the force on i-th particle from j-th particle 
is obtained as
\begin{eqnarray}
\nvec{F}_i  = 
     &- &\frac{   \nvec{r}_{ ij } }{r_{ij}} 
              \frac{ \partial V(r_{ i j})  }{ \partial r_{ i j} }
 \nonumber  \\     \  \     
    & - & 
     \frac{2}{r_{ij}} 
     \left(
          \nvec{m}^\perp_i  \cdot    \nvec{m}^\perp_j 
     \right)
    \left(
           m _{i}^\|  \nvec{m}^\perp_j
           +
           m _{j}^\|  \nvec{m}^\perp_i 
    \right)
   \label{model_force}
\end{eqnarray}   
using the gradient formula given in eq. \ref{transv_gradient}.

Now we address the question of the torque on the vectors  $ \nvec{m}_i $
due to their mutual interaction. While the above formulae are
equally valid both in two and three dimensions, we limit the discussion
here to two dimensions, since a full three-dimensional discussion
becomes somewhat lengthy. In two dimensions, there is only one angle
specifying the orientation of each of the vectors. Denoting $\alpha_i$ 
the orientation angle of i-th internal vector $\nvec{m}_{i}$ and
$\theta_{ij}$ the direction of vector $\nvec{r}_{ ij }$,
\begin{equation}
  \nvec{m}^\perp_i  \cdot    \nvec{m}^\perp_j 
  = \cos(\alpha_i -  \alpha_j) 
  - \cos (  \alpha_{ i  } -\theta_{ij} )   \   \cos (\alpha_{ j  }-\theta_{ij}  ) 
   \label{def_angles}
\end{equation}   
following directly (also valid in three dimensions) from
\begin{equation}
  \nvec{m}^\perp_i  \cdot    \nvec{m}^\perp_j 
  = 
   \nvec{m} _{ i}    \cdot    \nvec{m}_{ j} 
  -
  m _{ i}^\| m _{j}^\| 
   \label{def_scalar}
\end{equation}   
To evaluate the torque on $\nvec{m}_{ i}$ we need to find the
gradient (or derivative in this case)
\begin{equation}
   \frac{ \partial   }{ \partial \alpha_{ i} }
   \left(
  \nvec{m}^\perp_i  \cdot    \nvec{m}^\perp_j       \right)
 =
  \cos (  \alpha_{ i  } -\theta_{ij} )   
  \   \sin (\alpha_{ j  }-\theta_{ij}  ) 
   \label{def_torque}
\end{equation}   
  The torque on the $\nvec{m} _{ i}$ is zero when
  $ \alpha_i = \theta_{ij}  +\frac{  \pi }{ 2 }$, i.e. when it already is 
  perpendicular to the vector $\nvec{r}_{ ij }$, connecting the two particles.
In three dimensions the internal vectors are given by two angles, as well as the connecting
position vector, but the principle remains the same and the evaluation is  more complicated
but straightforward.

Possible applications of this model with fictive variables
must solve the problem of removing the energy associated 
with the fictive degrees of freedom. In our application
described elsewhere \cite{obmd} we use a model of 
over-damped angular motion of the internal vectors.
The excitation energy resulting from the nonoptimal orientation
is spurious, since the TDs are not real quantities. When the optimal
orienatation is reached, in normal Newtonian treatment
the potential energy would be turned into some (mostly rotational) 
kinetic energy, leading to spurious vibrations in the angles.
Thus, it should be a part of the model to dispose off this energy,
through a fictive friction. We propose that this can be done 
by considering an over-damped motion of the model transvers dipoles,
leading to first order equations, as described in detail in ref. \cite{obmd}.

It might be of interest to consider a further generalization of the discussed model,
by including both parts of the dipole interaction, as 
\begin{equation}
 W(r_{ij}) 
 =  
   t (r_{ij}) 
                \nvec{m}_{i}^\perp \cdot \nvec{m}_{j}^\perp 
   + 
   p (r_{ij})
                \nvec{m}_{i}^\|    \cdot \nvec{m}_{j}^\|
   \label{dipolegeneral}
\end{equation}   
Here the radial functions might be such that they would lead to parallel layers,
i.e. the $ p (r_{ij}) $ should have a repulsive part for small distances and a minimum
at the desired distance between two layers. The part $ t (r_{ij}) $ should be similar 
to the  $g(r_{ij})$ of eq. 
  \ref{dipoleeffective}, or somewhat modified to model even further features.

% The saturation feature
% We are preparing a paper
An extensive discussion of the applications of the
proposed  type of model interactions is discussed in our
above quoted paper \cite{obmd}. We would like to publish this 
shortened discussion of the model in this journal, 
since it might be useful in entirely different fields of research,
or at least inspire   similar applications, much in the same way as the
surprising model of isotropic pair interactions 
leading to complex space patterns of diamond-like lattice 
discussed above (ref. \cite{stillinger2007}) inspired our work.

%%%%%%%%%%%%%%%%%%%%%%%%%%%%%%%%%%%%%%%%%%%%%%%%%%%%%%%%%%%%%%%%%%%%%%%%%%%%%%%%%%%%
\section*{References}
%%%%%%%%%%%%%%%%%%%%%%%%%%%%%%%%%%%%%%%%%%%%%%%%%%%%%%%%%%%%%%%%%%%%%%%%%%%%%%%%%%%%%

\end{document}